\documentclass{optica-article}

\journal{opticajournal} 

\articletype{Research Article}

\usepackage[utf8]{inputenc}
\usepackage{float}
\usepackage{lineno}

\begin{document}

\title{Point Spread Function Engineering Using Implicit Neural Representations}

\author{Suet Ying Chan,\authormark{1*}, Mitchell Gilmore\authormark{1}, Qilin Deng\authormark{1}, Guorong Hu\authormark{1}, Joseph Greene \authormark{1, 2}, Ruipeng Guo\authormark{1}, Lei Tian\authormark{1}}

\address{\authormark{1} Boston University, Department of Electrical and Computer Engineering, Boston, MA, 02215}
\address{\authormark{2}Georgia Tech Research Institute, Electro-Optical Systems Lab, Atlanta, GA, 30332}

\email{\authormark{*}syc@bu.edu} 


\begin{abstract*} 
Point spread function (PSF) engineering through pupil plane modulation is a technique used in microscopy to achieve specific imaging properties, such as depth encoding or extended depth of field.
Existing PSF design methods often rely on extensive domain knowledge and task-specific basis functions, making it difficult to generalize across different applications.
We treat the PSF engineering task as a phase retrieval problem and propose a neural field pupil design method that optimizes a phase profile for any arbitrary, user-defined 3D PSF distribution.
This provides a flexible framework for 3D PSF engineering for various applications with implicit regularization that proves robust to initialization compared to pixel-wise optimization methods.
\end{abstract*}

\section{Introduction}
Wavefront engineering using pupil plane modulation has been widely used to design point spread functions (PSF) for various applications in imaging.
The diffraction-limited Airy disk PSF is suboptimal for many imaging tasks that require 3D information, as it does not encode depth information effectively.
Classical examples of controlling the depth-variance (or invariance) of the PSF through pupil engineering include the helical, extended-depth-of field (EDOF), and multifoci PSFs, each of which requires application-specific design approaches.

One prominent application of the helical PSF is in 3D localization microscopy, which aims to encode the precise 3D position for a point emitter to nanometer precision, and they are designed either from statistical criteria such as minimizing the Cram\'er--Rao lower bound (CRLB) \cite{Shechtmanetal2014} or from beam-shaping bases that directly shape the complex field, e.g. Laguerre--Gaussian modes \cite{piestun2002synthesis, pavani2008rotating}.
EDOF PSFs, on the other hand, relies on an invariant PSF through depth. These PSFs allow
for single-shot imaging of a 3D volume without the need for axial scanning, and their design is
traditionally accomplished using the cubic phase plate or the radial bases \cite{george2003extended, khonina2021axicons}.
On a similar note, in multifocal imaging, only select axial planes are excited rather than a continuous axial
range. Achieving precise control in multifocal excitation has various applications in intraocular
lenses, multiplane optical trapping, direct laser writing for microfabrication, etc. \cite{doskolovich2018multifocal, machado2018walsh, feng2023multifocal, zhu2021laser}, and they utilize different sets of radial bases, such as the Fresnel and Walsh zone plates, which requires a careful understanding of diffractive optics.
Each task demands understanding of different bases, and the precision and scale required vary depending on the specific application.
Choosing the correct basis a priori becomes a substantial part of the design problem. This makes it difficult to generalize the use of a-designed mask or design method across different applications.

An alternative is to treat the 3D PSF engineering task as a phase retrieval problem, which is analogous to the inverse problem in computer-generated holography (CGH) \cite{lee1978computer}, where the goal is to retrieve the phase profile that produces a desired 3D intensity patterns.
Phase retrieval is a well-studied technique in optics that recovers the phase information of a signal from intensity-only measurements \cite{fienup1982comparison, shechtman2015phase}.
There are many existing approaches, such as the Gerchberg-Saxton (GS) algorithm and its variants \cite{sui2024nonconvex}, which iteratively project between the spatial and Fourier domains to solve for the phase of the given intensity volume.
Gradient-based pixel-wise optimization methods have also been used to directly optimize for the phase values at each pixel on a discretized pupil grid \cite{zhang2017nonconvex, chakravarthula2019wirtinger}.
The bridge between inverse design and phase retrieval can be seen in the use of aberration estimation, where the non-ideal wavefront is retrieved given the aberrated PSF stack \cite{Paxmanetal1992}.
With the rise of deep learning, there has also been extensive work in parameterizing the forward wave optics model and letting neural networks learn the mappings from target image intensity to the phase \cite{neural3dholo2021}.

However, these pixel-wise phase retrieval optimizations are tailored to CGH applications and do not explore the need for micron-scale accuracy and the interpretability typically expected from the PSF engineering community.
There are also long-existing problems the community has been trying to solve, such as the optimization method being susceptible to stagnation (i.e., a local minimum solution) and high frequency speckle noise in the learned phase due to the lack of continuity between pixels \cite{sui2024nonconvex,zhang2017nonconvex}.
In traversing the non-convex loss landscape, this makes the optimization sensitive to initialization, and non-physically intuitive phase profiles are learned as a result of the pixel-wise optimization approach.
Therefore, despite these recent advances, stable optimization methods and physically interpretable phase representations for 3D PSF engineering remain underexplored.

To address this gap, we propose a neural field pupil design method for 3D PSF engineering.
It leverages implicit neural representations (INRs) \cite{mildenhall2020nerf} to parameterize the pupil phase mask using a multi-layer perceptron (MLP), which provides implicit regularization and removes the need to define a basis \emph{a priori}.
This allows for a smooth, continuous representation of the phase mask that is amenable to gradient-based optimization and robust to initialization compared to pixel-wise optimization methods.
Additionally, we incorporate a preprocessing step for the target PSF stack that mimics the band-limited frequency response of the optical system to ease convergence during training, which is tailored to PSF engineering and typically not considered in CGH applications.
The MLP is trained using a differentiable Fourier optics model.
Since the MLP learns a continuous functional representation, it can be queried at arbitrary spatial resolution without interpolation or retraining.
All together, the proposed method provides a flexible framework for 3D PSF engineering that is generalizable to various applications. This reduces the need for extensive domain knowledge for a given PSF design task and enables the user to specify arbitrary 3D PSF distributions for a given task.
We summarize results demonstrating various PSF designs and their experimental validation using a spatial light modulator (SLM), and demonstrate the advantages of the proposed method over state-of-the-art pixel-based optimization approaches.

\section{Methods}
\subsection{Neural Field Pupil Representation}
\begin{figure*}[htb]
\centerline{\includegraphics[width=0.8\linewidth]{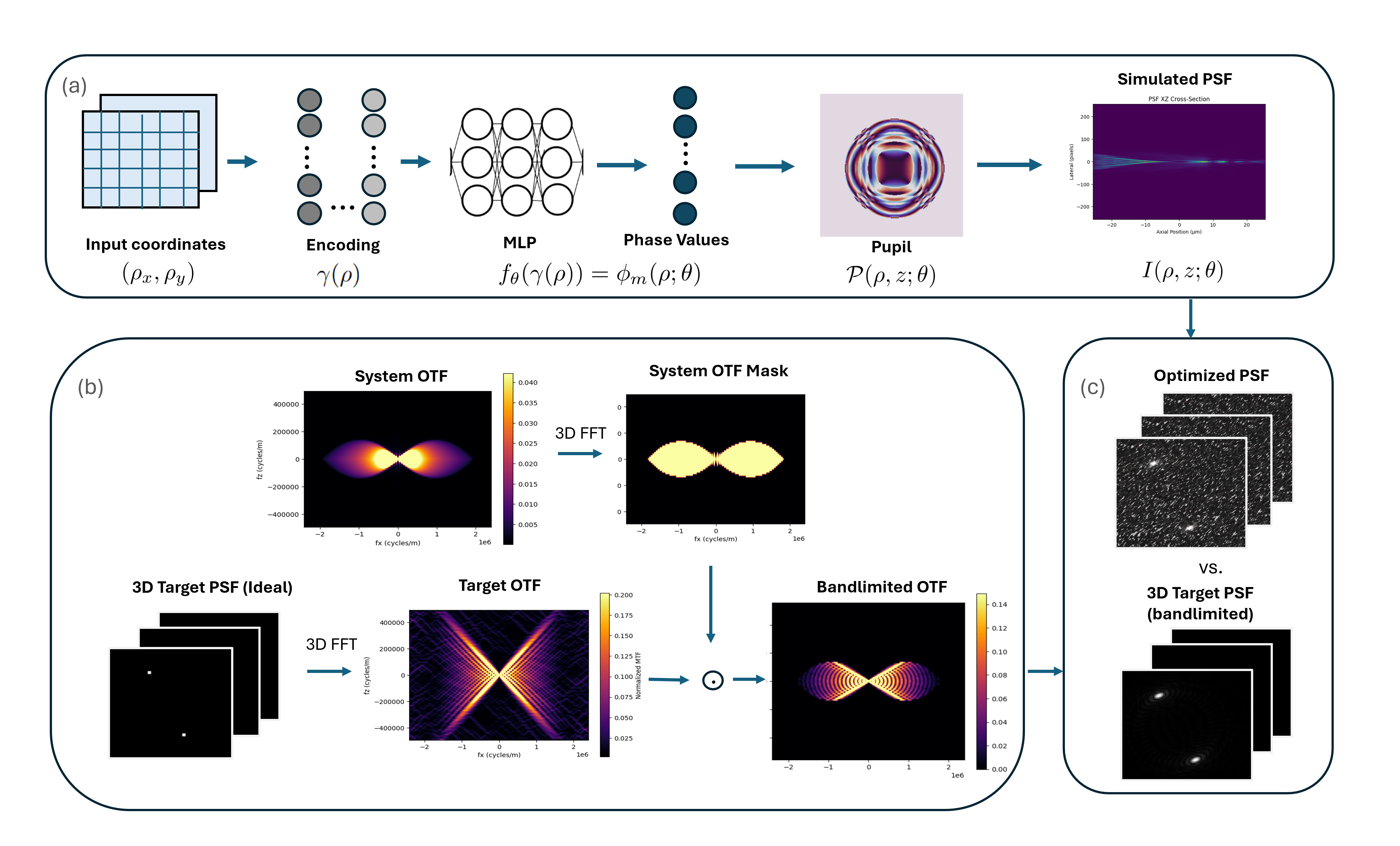}}
\caption{Overview of neural field pupil engineering pipeline. (a) Input normalized pupil coordinates are encoded using Fourier feature mapping and passed through an MLP to predict the phase value at those coordinates, which are remapped to the 2D pupil function to compute the 3D PSF stack. (b) The target PSF stack is band-limited by the system OTF. (c) The loss is computed between the predicted and target PSF stacks, and the MLP is optimized using a gradient-based algorithm.\label{fig:overview}}
\end{figure*}

The overall framework for the proposed neural field pupil engineering method is illustrated in Figure~\ref{fig:overview}.
Instead of using a fixed basis or pixel-wise representation for the phase mask, the pupil is parameterized using a multilayer perceptron (MLP) neural network. This coordinate-based model, also known as an implicit neural representation (INR), has been widely used in computer vision and graphics for representing signals \cite{mildenhall2020nerf, sitzmann2020siren}.
It has also been used in the optics community for applications such as aberration estimation \cite{feng2023neuws} and parametrization of the pupil function in solving inverse problems \cite{zhang2025wholefield}.
Unlike classical approaches that solve for the coefficients of a chosen basis \cite{Tseng2021NeuralNanoOptics}, our pipeline maps a normalized pupil coordinate to the phase at that coordinate.
Coordinate-based
representations are inherently continuous and treat images as functions (not matrices), thus
they do not restrict reconstruction to any subspace; i.e., for the task of phase mask optimization,
it does not require defining the basis a priori.

The INR optimization optimizes $f_{\theta}$, an MLP parameterized by $\theta$. The phase mask is modeled as:
\begin{equation}
    \phi_m (\rho; \theta) = f_{\theta}(\gamma(\rho))
    \label{eq:MLP}
\end{equation}

where $f_{\theta}: \mathbb{R}^{4L} \rightarrow \mathbb{R}$ is the MLP with learnable parameters $\theta$, $\gamma(\rho) : \mathbb{R}^{2} \rightarrow \mathbb{R}^{4L}$ is the Fourier feature mapping, and $\rho \in [-\frac{1}{2}, \frac{1}{2})$ are the normalized neural field coordinates.

Despite the benefits of implicit regularization, it also causes spectral bias, which means MLPs tend to learn low-frequency functions more easily than high-frequency ones \cite{mildenhall2020nerf, sitzmann2020siren}. This can be an issue when trying
to learn complex phase masks that require high-frequency variations. To overcome this,
the input normalized pupil coordinates are encoded using Fourier feature mapping, which
projects the coordinates into higher-dimensional space and is defined by a sinusoidal basis $
\gamma(\rho) = [\cos(2^0 \pi \rho), \ldots, \cos(2^{L-1} \pi \rho), \sin(2^0 \pi \rho), \ldots, \sin(2^{L-1} \pi \rho)]^T$. This gives a total of $2 \times L \times 2 = 4L$ features.

Note that directly learning the real-valued phase $\phi_m \in \mathbb{R}$ has an inherent phase wrapping ambiguity, as the complex exponential is invariant to $2\pi$ phase shifts, i.e., $\phi_m \mapsto \phi_m + 2\pi k$ for $k \in \mathbb{Z}$, which is a well-known phase-unwrapping problem \cite{kamilov2015phase}.
We therefore have the MLP learn the phase in cartesian form, with its two output channels corresponding to the real and imaginary components of the complex exponential representation of $e^{\phi_m}$, and normalize them to unit magnitude.
The phase mask is recovered using \begin{equation}
e^{i\phi_m(\rho;\theta)} =
\frac{f_\theta(\gamma(\rho))_1 + i\,f_\theta(\gamma(\rho))_2}
     {\left\| f_\theta(\gamma(\rho)) \right\|},
\label{eq:cartesian}
\end{equation}, where $f_{\theta}(\gamma(\rho))_1$ and $f_{\theta}(\gamma(\rho))_2$ are the first and second channels of the MLP output, respectively.
This ensures the output lies on the unit circle and the learned mask is phase-only (with amplitude assumed to be 1). The encoded coordinates are passed through the MLP with 3 layers and 256 hidden units. The learned
phase values are remapped to the 2D pupil function and used to compute the 3D PSF stack.
Another method of addressing the phase ambiguity using equivariance regularization is discussed in the Supporting Information.

The neural field pupil model is differentiable with respect to $\theta$, and gradient-based algorithms (e.g., Adam, L-BFGS) are used to minimize a task-specific loss.
Given a target PSF stack $I_t(x, y; z)$, the training objective is then:
\begin{equation}
\theta^* = \arg\min_{\theta} \mathcal{L}(\left\lvert \mathcal{F} \{\mathcal{P}_{\theta}(\hat \rho, z)\} \right\rvert^2, I_t(x, y; z)),
\end{equation}
where $\mathcal{L}$ defines the loss function.

\subsection{Fourier Optics Forward Model}
To compute the scalar field at some axial distance $z$ from a known field in the input plane, we use the angular spectrum method. Let $U(x, y)$ be the complex field distribution at $z = 0$, and let $\tilde U(f_x, f_y) = \mathcal{F}\{U(x, y)\}$ be its spatial frequency spectrum.
The field at distance $z$ is given by:
\begin{equation}
    E(x,y;z) = \mathcal{F}\left\{\tilde{U}(f_x,f_y)\, H_z(f_x,f_y)\right\},
\end{equation}
where the free-space transfer function is
$
    H_z(f_x,f_y) = e^{-i2\pi\frac{z}{\lambda}
    \sqrt{1-(\lambda f_x)^2 - (\lambda f_y)^2}}.
$ 
Given normalized pupil coordinates, $\rho = (\frac{fx \lambda}{NA}, \frac{fy \lambda}{NA})$, this can be written as the defocus kernel $\phi_d(\tilde{\rho}, z) = kz\sqrt{1-({\mathrm{NA}}/{n})^2\|{\hat \rho}\|^2}$ with
$k = 2\pi/\lambda$. The pupil function is then expressed as the point-wise multiplication of the neural field-learned mask and the defocus kernel:
\begin{equation}
    \mathcal{P}_{\theta}(\tilde{\rho}, z) = A(\tilde{\rho})\,
    e^{i\phi_m(\rho) + i\phi_d(\tilde{\rho}, z)},
    \label{eq:pupil}
\end{equation}
where $A(\tilde{\rho}) = \mathbf{1}$ for $\|\tilde{\rho}\| \leq 1$ is the circular aperture that defines the frequency cutoff in the normalized pupil grid.

The intensity PSF at depth $z$ is expressed as
\begin{equation}
    h(x,y;z) = \left|\mathcal{F}\left\{\mathcal{P}_{\theta}(\hat \rho;z)
    \right\}\right|^2.
    \label{eq:psf}
\end{equation}
The network is queried on normalized coordinates $\rho = n/N$, where $n = [-\frac{N}{2}, ... \frac{N}{2} -1]$. This fixes the domain and changes the spacing $\Delta \rho$ based on the array size. These coordinates correspond to the pupil coordinates through $\hat \rho = \frac{\rho \lambda}{\mathrm{NA} \Delta x}$, where $\Delta x$ is the object-space pixel pitch. The scaling contains no N, so for fixed system parameters ($\lambda$, NA, and $\Delta x$), a query at $\hat \rho$ addresses the same pupil position irrespective of array size. At deployment the mask is evaluated on a larger grid $N' > N$ to match the SLM resolution by re-evaluating $f_{\theta}$ and the encoding at the new coordinates, rather than resampling the trained $N \times N$ array. 

\subsection{Target 3D PSF Stack Generation}

For the purpose of PSF engineering, the target consists of mostly zeros (i.e., sparse), which can make the optimization difficult as the loss function is dominated by the background pixels.
Given the diffraction limited optical transfer function (OTF), which expresses the frequency response of the optical system \cite{goodman2017fourier}, we include a preprocessing step on the target PSF stacks to mimic its band-limited nature.
To generate the target PSF stacks for training, the user can specify a set of delta functions at the desired location at each axial slice, $h_{\mathrm{ideal}}(x,y,z)=\sum\limits_{k=1}^{K}\delta(x - x_k,\, y - y_k)\,\delta(z - z_k).$
A binary support mask ($M_{\mathrm{OTF}}$) is obtained by thresholding the magnitude of the 3D OTF, computed as the Fourier transform of the simulated diffraction-limited PSF stack for the given optical parameters (e.g., NA, wavelength, pixel size, etc.).
The filtered target spectrum is
\begin{equation}
    \tilde{H}_{\mathrm{target}}(f_x,f_y,f_z) =
    M_{\mathrm{OTF}}(f_x,f_y,f_z)\,
    \tilde{H}_{\mathrm{ideal}}(f_x,f_y,f_z),
\end{equation}
and the final target PSF stack is recovered as
$I_t(x,y;z) = \mathcal{F}^{-1}\{\tilde{H}_{\mathrm{target}}\}$.

Each axial slice of the target is normalized to unit total intensity, matching the normalization of the learned PSF in Equation~(\ref{eq:psf}).
Since the learned mask is phase-only, $|\mathcal{P}_\theta|$ is independent of
$z$. By Parseval's theorem, every axial slice of the computed stack carries
the same total intensity. Without per-slice normalization, a plane containing more target points would be assigned more energy than planes with fewer target points.
The relative intensity between slices is then reintroduced through a user-defined weight $\alpha_k$ for each slice $z_k$, which lets the optimizer prioritize focus in the higher weighted slices. This can be useful to compensate intensity falloff from scattering and absorption in 3D imaging applications.
Thus, the intensity weighted objective function can be defined as
\begin{equation}
\mathcal{L} = \sum_{k} \alpha_{k}\,
\bigl\| I_t(x,y;z_k) - h(x,y;z_k;\theta) \bigr\|^{2},
\label{eq:loss}
\end{equation}
where $I_t(x,y;z_k)$ is the target PSF slice at axial position $z_k$, $h(x,y;z_k)$ is the predicted PSF slice from Equation~(\ref{eq:psf}), and $\alpha_k$ is the user-defined weight factor for that slice.

\begin{figure}[htb]
\centering\includegraphics[width=0.8\columnwidth]{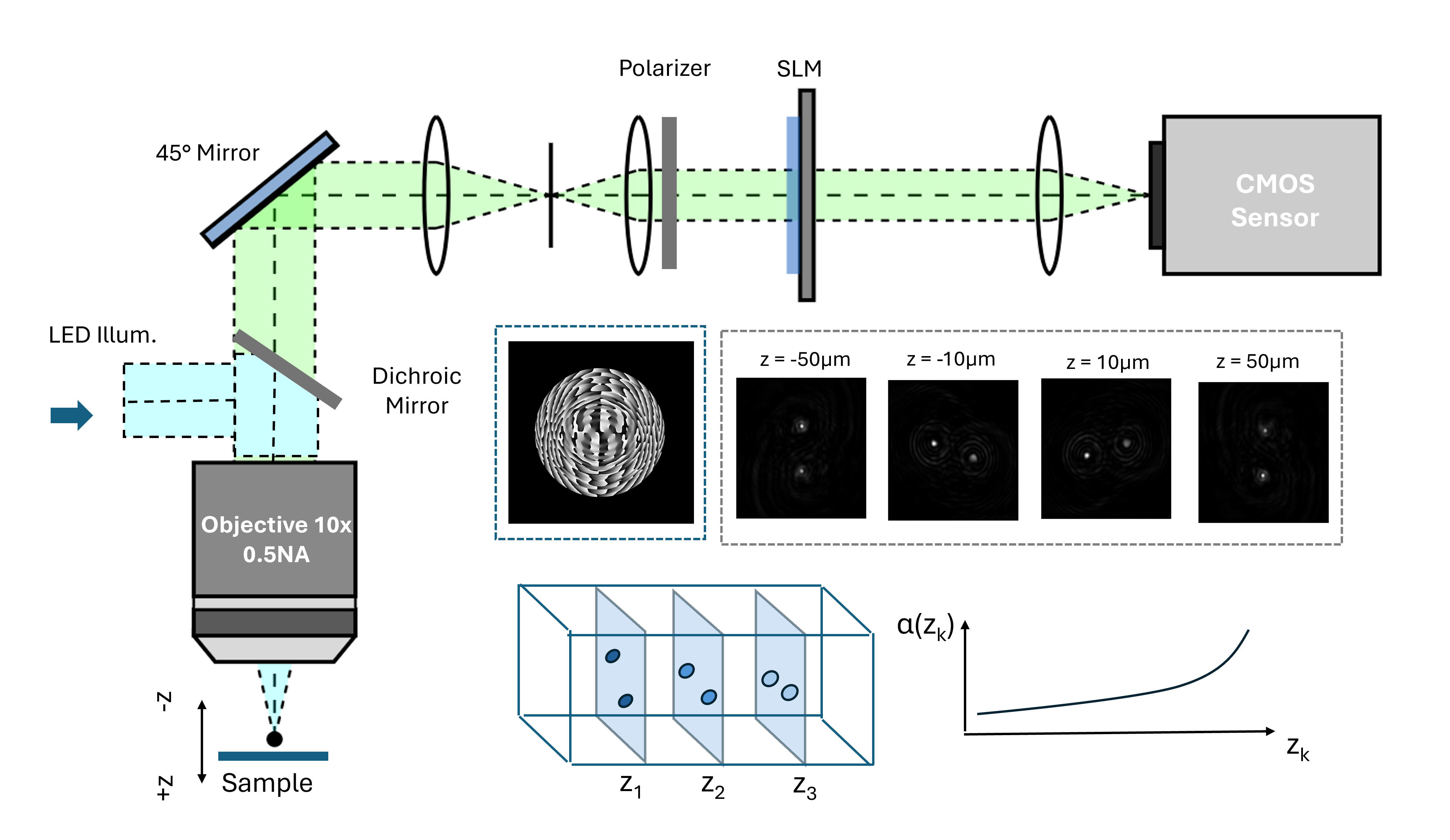}
\caption{Diagram of tabletop optical system with the spatial light modulator (SLM). Learned phase masks are loaded onto the SLM, and the 3D PSF stack are obtained experimentally by axial scanning of a fluorescent bead. The relative peak intensity of the 2D PSF between the axial slices can be controlled by the user-defined weight factors for each axial slice.\label{fig:sys_diagram}}
\end{figure}

\subsection{Experimental Setup}

The illumination source is a 470\,nm LED (SOLIS-470C, Thorlabs) for fluorescence
excitation. Fluorescent beads (1\,$\mu$m diameter) were used as point-source
for PSF collection. The emitted light is passed through a dichroic mirror (Thorlabs, MDF05-GFP).
The Thorlabs Kinesis motorized stage is used for automatic axial scanning of the sample stage.
Emitted fluorescence is collected by a 10$\times$ objective (CFI Super Fluor 10$\times$,
NA\,=\,0.5, WD\,=\,1.1\,mm; Nikon) and a tube lens (TTL200-A, Thorlabs). The beam is directed onto a reflective liquid-crystal-on-silicon spatial light modulator (1920 × 1152 XY Phase Series, HSP1920; Meadowlark Optics, 9.2 µm pixel pitch). The system is designed to demagnify the objective lens pupil onto the active area of the SLM, which is mounted onto a 3-axis stage (Thorlabs PT3A). The modulated beam is relayed onto a CMOS camera (DMK 33UX546;
2840$\times$2840 pixels, 2.74\,$\mu$m pixel pitch). The magnification after the tube
lens is 1.33$\times$, yielding a total system magnification of 13.3$\times$ from sample to camera.
Prior to loading onto the SLM, the neural field output is phase-wrapped and converted to an 8-bit grayscale value. SLM calibration was performed at the fluorescence central emission wavelength (520 nm) using
the vendor-supplied LUT calibration software. An additional sensorless wavefront correction
step \cite{booth2007sensorless} can be applied to compensate for system-level aberrations; the resulting correction phase
can be superimposed on the learned mask at deployment to improve PSF fidelity.
During inference of the neural field, the scale factor to map the training grid to the SLM display resolution is adjusted so the mask fills the pupil.

\section{Results}\label{sec3}

\begin{figure}[htb]
\centering\includegraphics[width=0.8\columnwidth]{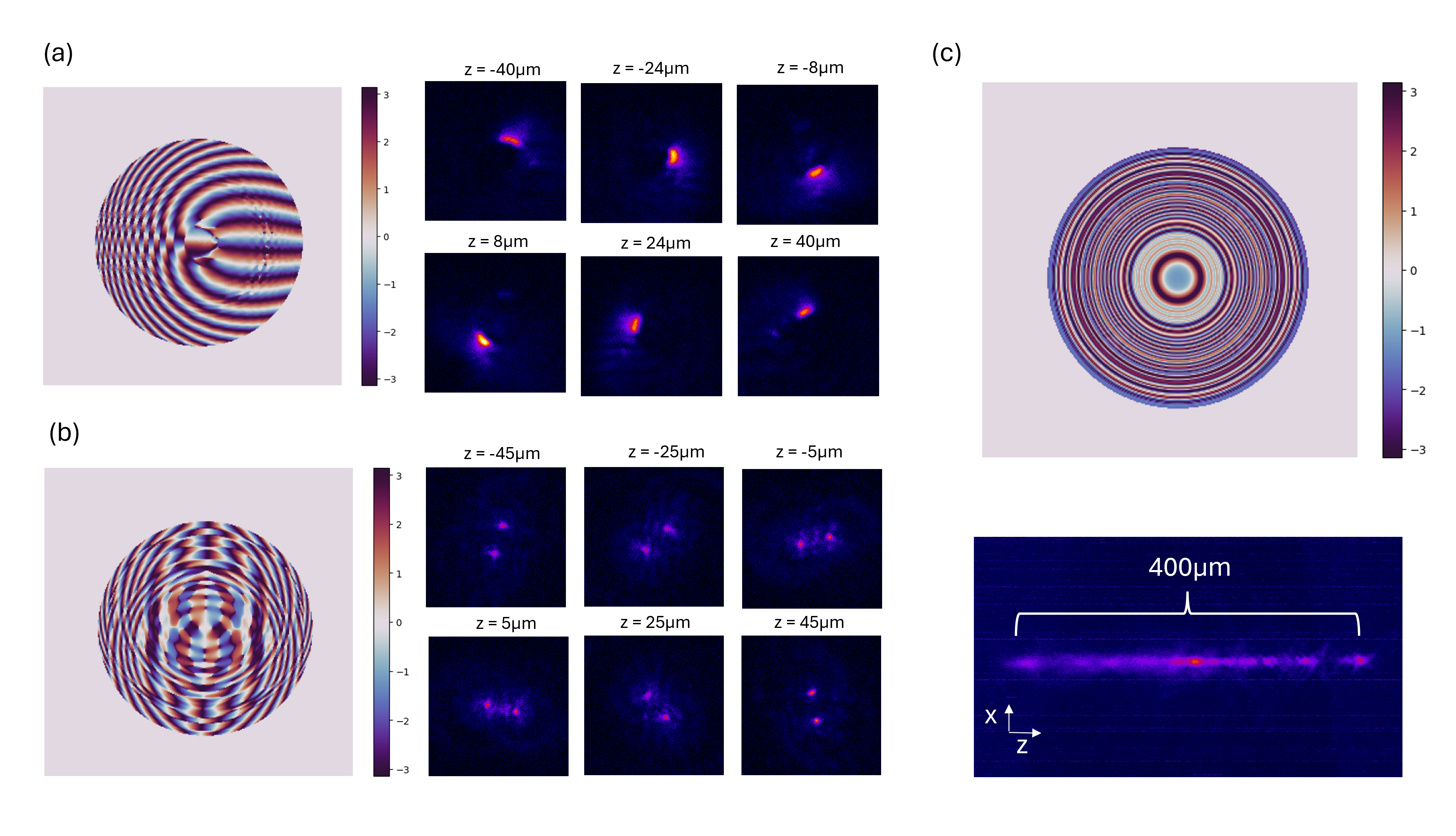}
\caption{Experimental results showing the lateral PSF at select depth slices for the (a) single helix PSF (designed for 2$\pi$ rotation over 80 $\mu$m axial range and (b) double helix PSF (designed for $\pi$ rotation over 100$\mu$m axial range), and (c) EDOF PSF (designed for 400 $\mu$m axial range).\label{fig:exp_results_1}}
\end{figure}

Figure~\ref{fig:exp_results_1} shows the experimental results for well-known, classical PSFs. The single-helix and double-helix PSFs were designed to rotate \(360^\circ\) and \(180^\circ\) through axial ranges of \(80~\mu\mathrm{m}\) and \(100~\mu\mathrm{m}\), respectively. The results show that the INR-optimized masks can learn this desired PSF behavior and are effective when deployed in a tabletop system.
The axial extension and rotation of the PSFs matched well with the simulation results.
The EDOF PSF was designed to be invariant through an axial range of \(400~\mu\mathrm{m}\), and the results also show that the PSF maintains extension through the designed axial range.

\begin{figure}[htb]
\centering\includegraphics[width=0.8\columnwidth]{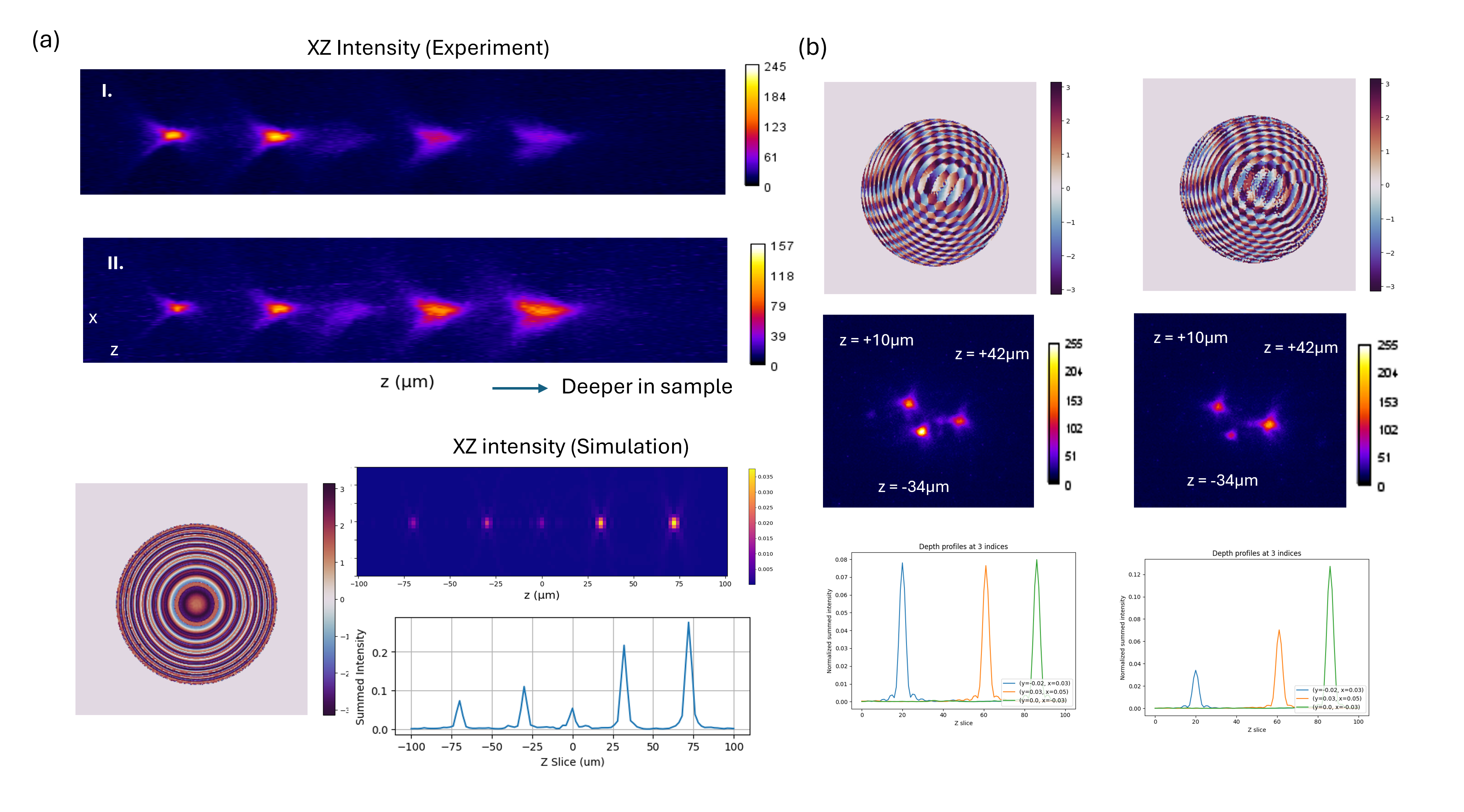}
\caption{Experimental results demonstrating the effect of weighing relative intensities between target points. (a) Multifoci results (i) without explicit weighting and (ii) with explicit weighting to enforce more uniform intensity between focal points (corresponding simulation target and results shown below). (b) Arbitrary point excitation and weighing the farthest (+z) point to have a higher intensity than the closest (-z), again demonstrating intensity preservation at a deeper imaging depth.\label{fig:exp_results_2}}
\end{figure}

Control over the relative intensities between focal points can be useful in 3D imaging, where intensity falloff due to scattering and absorption can be compensated for by adjusting the relative intensities of the focal points in optimization.
Figure~\ref{fig:exp_results_2} shows experimental results for controlling the relative intensities between multiple focal points.
For Figure~\ref{fig:exp_results_2}(a), the target PSF was designed to counter the intensity falloff at farther depths. The target PSF was weighted to have uneven intensities between the focal points.
The axial slices at farther propagation distances (+$z$) were weighted with higher target intensity, which was effective in evening out the intensity drop-off through depth.

Figure~\ref{fig:exp_results_2}(b) demonstrates another example of relative intensity control for a random focal point distribution.
The target PSF stack was designed to have 3 focal points at random lateral and axial locations.
Left and right experimental results show the difference in relative intensities with and without explicit weighting.
This could be useful in applications where the focal points are distributed across a large axial range and intensity falloff is more severe at deeper planes, such as in biological issue.

\begin{figure}[htb]
\centering\includegraphics[width=0.8\columnwidth]{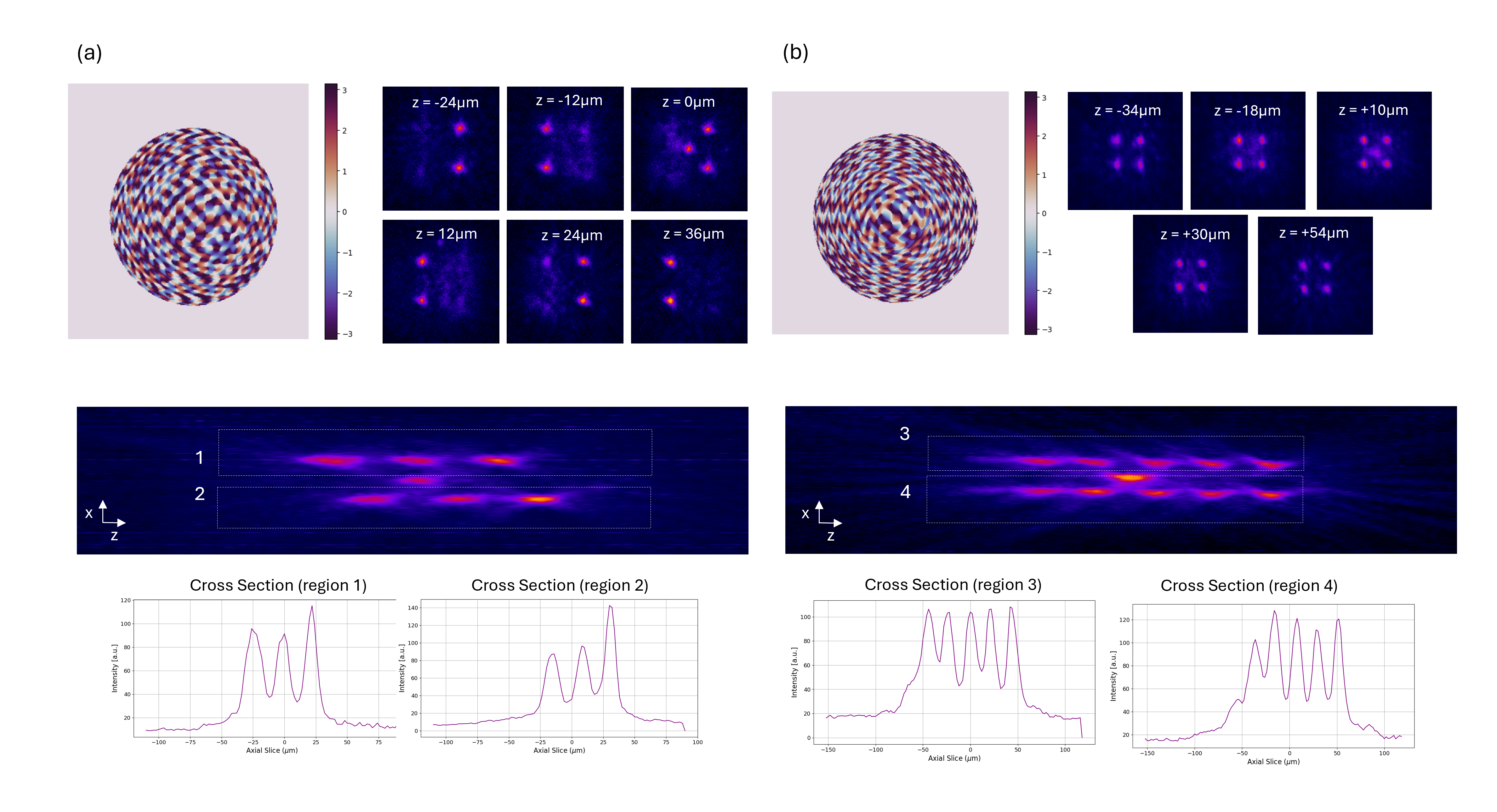}
\caption{Additional arbitrary PSF designs and experimental results for targeted illumination and multi-plane imaging, such as (a) an alternating 3-foci PSF with staggered axial spacing and (b) 2 $\times$ 2 multifoci PSF.\label{fig:exp_results_3}}
\end{figure}

\begin{figure}[htb]
\centering\includegraphics[width=0.6\columnwidth]{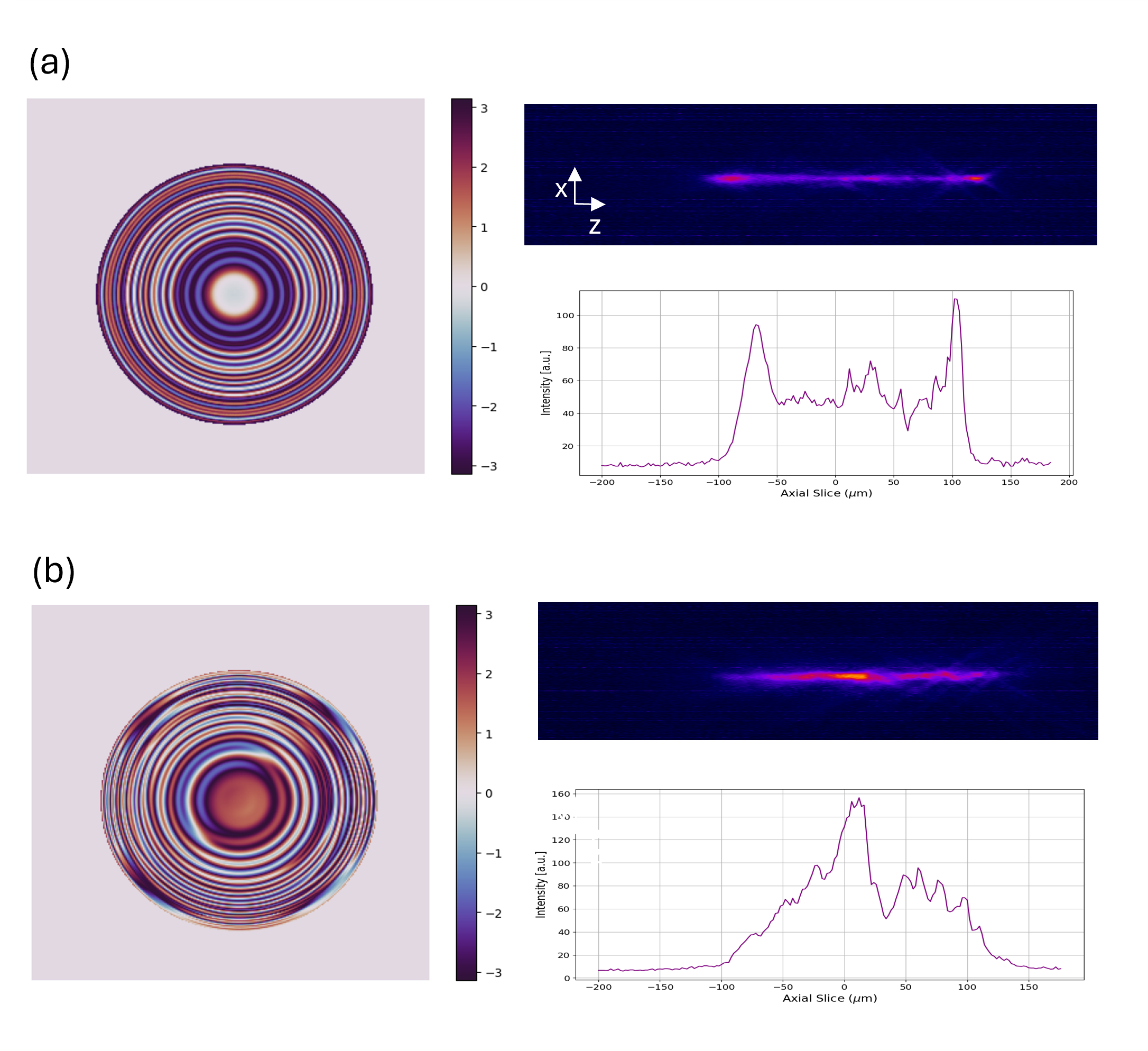}
\caption{Extended depth of field (EDOF) optimization and experimental results for 200 $\mu$m with (a) MSE loss and (b) MSE loss plus a regularization term to enforce greater concentration of energy within the central lobe.\label{fig:exp_results_4}}
\end{figure}

Figure~\ref{fig:exp_results_3} shows additional experimental results for arbitrary PSF designs, such as an alternating 3-foci PSF with staggered axial spacing and a 2$\times$2 multifoci PSF. Since the forward model assumes plane wave illumination at the pupil plane, the simulation pipeline can be easily translated to other general beam shaping purposes, such as targeted illumination and multi-plane imaging.
In addition, since this is a gradient-based optimization method, the loss function can be tailored with regularization terms to further tailor the optimization process.
For instance, Figure~\ref{fig:exp_results_4} shows the results for an EDOF design for an axial range of 200 $\mu$m with and without a regularization term that enforces greater concentration of energy within the central lobe.
The results show that the regularization term is effective and yields a phase mask design that breaks rotational symmetry, suggesting that that optimization can potentially allow degrees of freedom to counter non-ideal effects (e.g., higher order aberrations, scattering) in more sophisticated, end-to-end designs.
The details of the regularization term are included in the Supporting Information.
Furthermore, the axial range can also be easily extended to a larger range (e.g., 400 $\mu$m) by modifying the target PSF stack in the training objective, as demonstrated previously in Figure~\ref{fig:exp_results_1}.

\section{Discussion}\label{sec4}

\begin{figure}[htb]
\centering\includegraphics[width=\columnwidth]{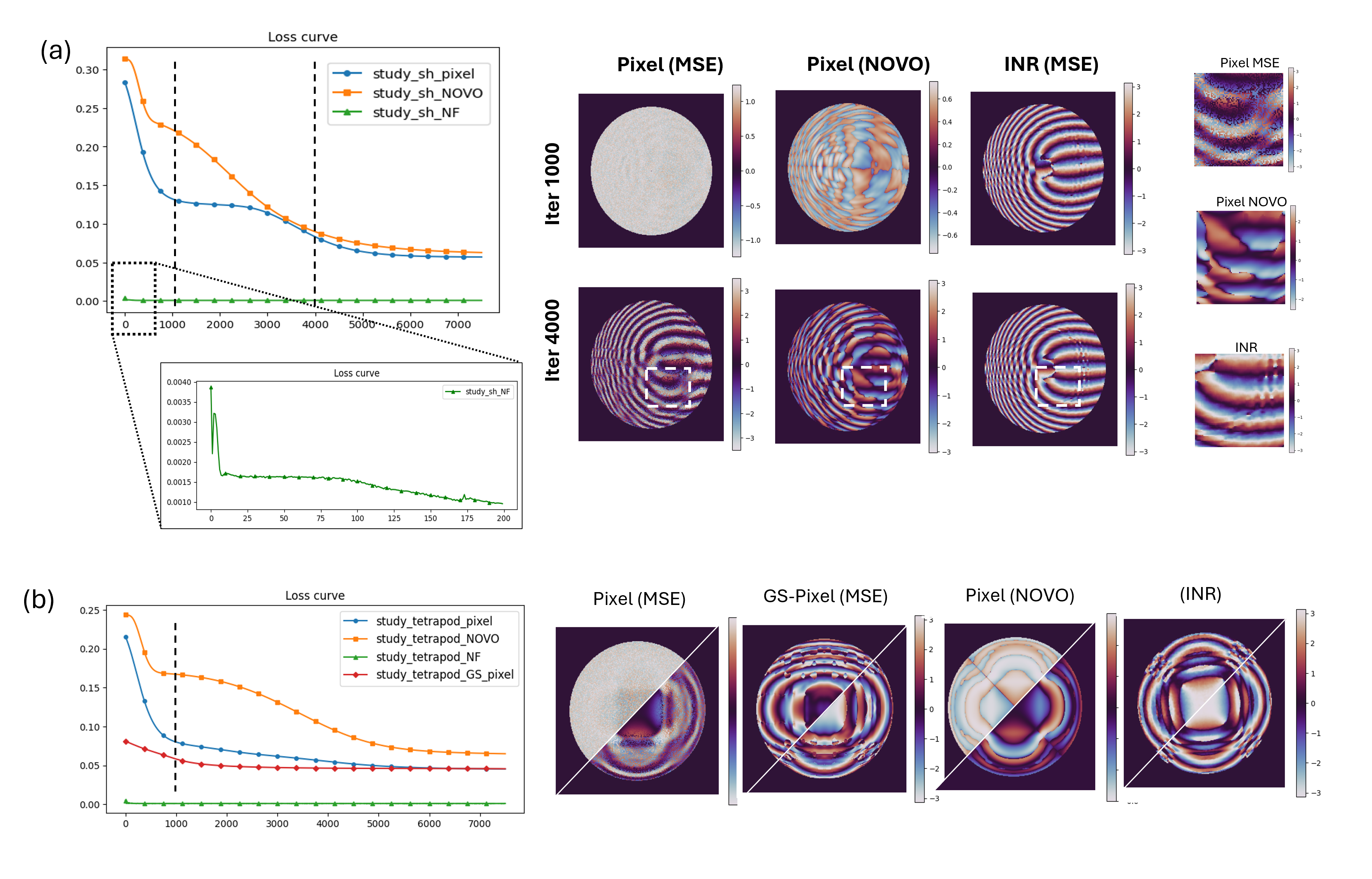}
\caption{Convergence results for different optimization methods. (a) The proposed neural field method converges with fewer iterations than pixel-wise optimization methods and produces a smoother phase profile for the single helix PSF. (b) Results for a tetrapod PSF design at epochs 1000 (upper left) and 4000 (lower right). Note all mthods were evaluated using MSE. \label{fig:comp_results_1}}
\end{figure}

The INR-optimized results were compared with other pixel-based optimization methods.
The pixel-wise optimization was implemented by directly optimizing for the phase values at each pixel in the discretized pupil grid.
PyTorch Autograd automatically handles the gradient computation through the complex field, so no custom gradient functions were needed, and the Adam optimizer was used for optimization.
The pixel-NOVO optimization was adapted from prior work \cite{zhang2017nonconvex}, where a custom cost function was designed to give the algorithm flexibility in finding suitable solutions for typically infeasible intensity distributions.
It specifies dark and bright regions via intensity thresholds. The details of the cost function is included in the Supporting Information.
Except for the NOVO optimization, all other optimization methods used the mean squared error (MSE) loss function between the predicted and target PSF stacks.

Figure~\ref{fig:comp_results_1} shows convergence results for the different optimization methods for the single-helix PSF design task.
The pixel-MSE optimization needed more iterations to find a solution without initialization, and the final learned phase profile was noisy due to the lack of regularization to enforce smoothness or continuity between the pixels.
High frequency speckle artifacts are common in pixel-wise optimizations in phase retrieval or CGH problems for this reason \cite{sui2024nonconvex}.
Avoiding high frequency noise in the learned mask is a desirable quality for imaging performance, especially in microscopy, as spurious hotspots or incorrect phase wrapping can lead to speckle noise when imaging and compound other issues in experimental conditions \cite{li2022speckle} (e.g., such as quantization error and interpolation of the trained mask to the SLM resolution).

The pixel-NOVO optimization needed more iterations to find a good solution, but it converged to a functional solution without the need for the GS initialization, confirming the utility in using a less stringent cost function in navigating the non-convex loss landscape.
However, the pixel-NOVO optimizations tend to have a higher loss than the MSE-based optimizations, which suggests that it trades off accuracy and convergence capability.
While CGH applications may not require a high degree of accuracy in the learned phase mask, PSF engineering applications typically require micron-scale accuracy.

Figure~\ref{fig:comp_results_1}(b) shows convergence results for the an example tetrapod PSF design.
Similar trends are observed for the loss curves.
GS-initialization (that initializes the phase with 50 iterations of the Gerchberg-Saxton algorithm) helps the pixel-wise optimization converge faster, but again, it is still susceptible to high frequency speckle noise in the learned phase profile.
The NOVO optimization converges without initialization, but it takes more iterations to converge and has a higher loss than the MSE-based optimizations.
Full results for the benchmarking study are shown in the Supporting Information.

\section{Conclusions}\label{sec5}
The proposed neural field pupil engineering method provides a flexible framework for 3D PSF engineering that is generalizable to various applications and makes task-specific PSF engineering more accessible.
Compared to other methods, our proposed method converged the fastest, did not require initialization, and did not struggle with high frequency speckle noise in the phase profile due to its inherent regularization.
The neural field method, an alternative to methods such as the NOVO loss, inherently learns a smooth, continuous phase profile that also has faster convergence.
The stability in training and flexibility in design can be extended to more sophisticated, end-to-end pipelines \cite{jin2020deep} to co-design the PSF for specific applications and experimental conditions, such as in in-vivo imaging where scattering and aberrations are more severe and need post-processing.
The compactness of the neural field representation also naturally extends high dimension inputs and is amenable to multi-wavelength \cite{shechtman2016multicolour} or spatially variant PSF designs \cite{xiao2024ppg3d}, which can be useful for applications such as multi-color imaging and wide field-of-view imaging.



\begin{backmatter}
\bmsection{Funding}
National Institutes of Health (R01NS126596), a grant from 5022 - Chan Zuckerberg Initiative DAF, an advised fund of Silicon Valley Community Foundation. 

\bmsection{Acknowledgment}
The authors acknowledge Boston University Shared Computing Cluster for proving the computational resources.

\bmsection{Disclosures}
The authors declare no conflicts of interest.

\bmsection{Data Availability Statement}
Data underlying the results presented in this paper and trained checkpoints to reproduce the results are not publicly available at
this time but may be obtained from the authors upon reasonable request.

\bmsection{Supplemental document}
A supplemental document must be called out in the back matter so that a link can be included. For example, “See Supplement 1 for supporting content.” Note that the Supplemental Document must also have a callout in the body of the paper.

\end{backmatter}



\bibliography{references}

\end{document}